\documentclass[twocolumn,aps,superscriptaddress]{revtex4-2}
\usepackage{hyperref}
\usepackage{xcolor}
\usepackage{physics}
\usepackage{amsmath}
\usepackage{amssymb}
\usepackage{comment}
\usepackage{algpseudocode}
\usepackage{subfigure}
\usepackage{graphicx}
\usepackage{geometry}

\usepackage[ruled,vlined]{algorithm2e}
\algrenewcommand\algorithmicrequire{\textbf{Input:}}
\algrenewcommand\algorithmicensure{\textbf{Output:}}
\usepackage{array,multirow}
\usepackage{rotating}

\begin{document}

\title{Reducing Measurement Costs by Recycling the Hessian in Adaptive Variational Quantum Algorithms}

\author{Mafalda Ramôa}
\email{mafalda@vt.edu}
\affiliation{Department of Physics, Virginia Tech, Blacksburg, VA, 24061, USA}
\affiliation{Virginia Tech Center for Quantum Information Science and Engineering, Blacksburg, VA 24061, USA}
\affiliation{International Iberian Nanotechnology Laboratory (INL), Portugal}
\affiliation{High-Assurance Software Laboratory (HASLab), Portugal}
\affiliation{Department of Computer Science, University of Minho, Portugal}

\author{Luis Paulo Santos}
\affiliation{International Iberian Nanotechnology Laboratory (INL), Portugal}
\affiliation{High-Assurance Software Laboratory (HASLab), Portugal}
\affiliation{Department of Computer Science, University of Minho, Portugal}

\author{Nicholas J. Mayhall}
\affiliation{Virginia Tech Center for Quantum Information Science and Engineering, Blacksburg, VA 24061, USA}
\affiliation{Department of Chemistry, Virginia Tech, Blacksburg, VA, 24061, USA}

\author{Edwin Barnes}
\affiliation{Department of Physics, Virginia Tech, Blacksburg, VA, 24061, USA}
\affiliation{Virginia Tech Center for Quantum Information Science and Engineering, Blacksburg, VA 24061, USA}

\author{Sophia E. Economou}
\affiliation{Department of Physics, Virginia Tech, Blacksburg, VA, 24061, USA}
\affiliation{Virginia Tech Center for Quantum Information Science and Engineering, Blacksburg, VA 24061, USA}

\begin{abstract}
Adaptive protocols enable the construction of more efficient state preparation circuits in variational quantum algorithms (VQAs) by utilizing data obtained from the quantum processor during the execution of the algorithm. This idea originated with ADAPT-VQE, an algorithm that iteratively grows the state preparation circuit operator by operator, with each new operator accompanied by a new variational parameter, and where all parameters acquired thus far are optimized in each iteration. In ADAPT-VQE and other adaptive VQAs that followed it, it has been shown that initializing parameters to their optimal values from the previous iteration speeds up convergence and avoids shallow local traps in the parameter landscape. However, no other data from the optimization performed at one iteration is carried over to the next. In this work, we propose an improved quasi-Newton optimization protocol specifically tailored to adaptive VQAs. The distinctive feature in our proposal is that approximate second derivatives of the cost function are recycled across iterations in addition to optimal parameter values. We implement a quasi-Newton optimizer where an approximation to the inverse Hessian matrix is continuously built and grown across the iterations of an adaptive VQA. The resulting algorithm has the flavor of a continuous optimization where the dimension of the search space is augmented when the gradient norm falls below a given threshold. We show that this inter-optimization exchange of second-order information leads the approximate Hessian in the state of the optimizer to be consistently closer to the exact Hessian. As a result, our method achieves a superlinear convergence rate even in situations where the typical implementation of a quasi-Newton optimizer converges only linearly. Our protocol decreases the measurement costs in implementing adaptive VQAs on quantum hardware as well as the runtime of their classical simulation.
\end{abstract}

\maketitle

\section{Introduction}

Despite the limitations of current quantum devices, there is still enormous interest in seeing whether they can provide quantum advantage in the simulation of strongly correlated quantum many-body systems. In particular, the variational quantum eigensolver (VQE) \cite{Peruzzo2014} was proposed as a near-term algorithm for many-body fermionic problems. In contrast with algorithms designed for the fault-tolerant quantum computing era, VQE employs shallow circuits and undertakes a naturally noise-resilient learning strategy, where a classical optimizer is used to tune parameters of quantum gates.

A crucial part of VQE is the \textit{ansatz}: it impacts accuracy as well as near-term viability. Problem-agnostic options have been shown to induce trainability issues termed \textit{barren plateaus} \cite{McClean2018}. Additionally, the coherence time requirements of current ansätze are, for reasonable-sized problems, still beyond what today's quantum computers have to offer. As such, new approaches attempting to further decrease circuit depth and improve trainability have emerged.

One promising option is to build the ansatz adaptively, such that its structure is dictated by the problem at hand. The first algorithm to use such a strategy was the Adaptive Derivative-Assembled Problem-Tailored VQE (ADAPT-VQE), proposed in Ref.~\cite{Grimsley_2019} and extended in Refs.~\cite{Tang2021,yordanov2020,Anastasiou2022,shkolnikov2021,Bertels2022,long_2023}. Starting from a very simple reference state, ADAPT-VQE creates a circuit block-by-block using information available on the fly. In the simplest version, each iteration selects one anti-Hermitian operator from a pre-defined pool, based on the associated gradient magnitude. This operator is multiplied by a variational parameter and exponentiated to create a parameterized unitary operator, which is appended to the ansatz. A VQE subroutine is then employed to optimize all of the parameters. After a new addition, the initial parameter vector for the optimization is built by appending a zero to the final parameter vector from the previous optimization, such that the newly added unitary is initially the identity. ADAPT-VQE leads to shallow and problem-tailored ansätze, and its parameter value recycling strategy makes it resilient against local traps \cite{Grimsley2022}. 

Inspired by ADAPT-VQE, other variational quantum algorithms (VQAs) have since employed this idea of iteratively growing the ansatz circuit as well as the gradient vector. An example is ADAPT-QAOA, proposed in Ref.~\cite{Zhu2020} and extended/studied in Refs.~\cite{chen2023,sridhar2023,yanakiev2023}. Other algorithms employ a fixed ansatz structure, but still augment the parameter vector iteratively to alleviate local traps and help avoid barren plateaus \cite{skolik2020,lee2021}. These strategies can be thought of as a special case of ADAPT-VQE where the selection criterion is trivial (pre-determined operators) but the parameter recycling strategy is employed regardless. 

The choice of which classical optimizer to employ in a VQA is important, as it impacts the costs of the algorithm as well as the solution quality. BFGS, a quasi-Newton optimizer which uses approximate second derivatives of the cost function to inform the search direction while avoiding the (costly) requirement of explicitly measuring them \cite{Dennis_1977,Dennis_1996,Flet_87,num_optimization}, is a popular choice \cite{guerreschi_2017,Lotshaw_2021,Grimsley_2019,Yordanov_2021,mbeng_2019}. A typical implementation of this algorithm will initialize an approximate inverse Hessian $H$, a matrix with all second derivatives, at the identity. Curvature information is gathered as the iterations proceed by performing rank-2 updates on this matrix, which are based solely on changes in the gradient and parameter vectors. As $H$ is updated, the search direction evolves from a vanilla gradient descent into an approximation to Newton's direction.

In a typical implementation of an adaptive VQA with the BFGS optimizer, the approximate inverse Hessian is reinitialized to the identity matrix each time the ansatz is extended. The reinitialization of $H$ seems discordant with the recycling of the parameters: The initial state at iteration $n$ is exactly the same as the final state from iteration $n-1$, and thus all previous approximate second derivatives remain as accurate as before.

In this work, we propose a strategy to recycle the Hessian in adaptive VQAs. At the end of each VQE subroutine, we save the collected curvature information and use it to inform the initialization of $H$ for the next optimization. With our protocol, the state of the optimizer, with all its knowledge of the cost landscape, is transferred between optimizations instead of being erased and restarted from scratch. We show that as a result, the search direction aligns more quickly with Newton's direction, and superlinear convergence is attained more often. Surprisingly, recycling old second derivatives allows the optimizer to better approximate new ones too, allowing new correlations to be captured earlier in the optimization. Our strategy leads to a significant decrease in the number of function evaluations required per optimization, which translates to a decrease in the number of calls to the quantum processor. For the molecules we studied, with 12 and 14 qubits, our protocol reduces the total measurement cost of the algorithms by an order of magnitude. We expect the impact to become more significant for larger system sizes.

The measurement costs are one of the most critical limitations of adaptive VQAs. They may be alleviated in two ways: by decreasing the number of measurements required to obtain the pool gradients, or by decreasing the number of measurements required in the optimization. Previous works have focused on the former \cite{shkolnikov2021,anastasiou2023}, such that the cost bottleneck now resides in the optimization process. To our knowledge, the strategy we propose is the first to tackle this source of costs, and the first to propose an optimizer specifically tailored to adaptive VQAs. We note that its relevance is not limited to implementations on quantum hardware---it also reduces the runtime of classical simulations of these algorithms, which can take hours even for relatively small molecules such as H$_6$ on a minimal basis set (12 qubits).

This paper is organized as follows. In Sec.~\ref{s:background}, we cover the relevant background topics: ADAPT-VQE and gradient-based optimization methods. In Sec.~\ref{s:bfgs_for_adapt} we introduce our main proposal, a BFGS algorithm tailored to ADAPT-VQE. In Sec.~\ref{s:results}, we present results from numerical simulations proving that our method is capable of significantly reducing the total number of calls to the quantum computer required by ADAPT-VQE. We additionally examine quantitative markers along specific optimizations to justify the remarkable speed up in convergence, and seek to understand under which conditions the advantage is the greatest by analyzing the distance between the initial approximate inverse Hessian and the initial exact inverse Hessian, as well as the difference between the latter and the final exact inverse Hessian. We conclude in Sec.~\ref{s:conclusion}.

\section{Background}
\label{s:background}
\subsection{Variational Quantum Algorithms}

VQAs are hybrid quantum-classical algorithms whose goal is to minimize a cost function. This cost function is typically formulated as the expectation value of a quantum Hamiltonian $\mathcal{\hat{H}}$, which can be written as a linear combination of Pauli strings and measured in a quantum computer via sampling \cite{McClean2016}.

The role of the quantum computer is then to host trial states that can be measured to evaluate the cost function. Such states are prepared by a parameterized quantum circuit, the \textit{ansatz}, which is usually comprised of two parts: the preparation of a reference state $\ket{\psi_{ref}}$ and the application of a parameterized unitary $U(x)$. $x$ is a vector whose elements map to gate parameters (e.g., angles of single-qubit rotation gates). In the interest of clarity, we will always denote parameter vectors by $x$ and the individual entries by $\theta_i$, such that if our parameter vector is $n$-dimensional we have $x=\{\theta_1,...,\theta_{n}\}$. This allows us to clearly distinguish iterations of the numerical optimizer, where $x_k$ is used to denote the parameter vector at iteration $k$, from iterations of adaptive VQAs, where $\theta_i$ is used to denote the parameter added at iteration $i$.

For a given ansatz and cost function, a VQA will seek the state corresponding to the lowest value of the cost function within the search space defined by the ansatz. This minimization is done by a classical optimizer.

The variational quantum eigensolver (VQE) is a subclass of VQAs aimed at finding eigenstates and eigenvalues of physical systems \cite{Peruzzo2014}. We focus on quantum chemistry and discuss VQEs for finding molecular ground states under the Born-Oppenheimer approximation. 

The first ansatz proposed for this problem was the Unitary Coupled Cluster Singles and Doubles (UCCSD) ansatz, motivated by classical variational methods for quantum chemistry \cite{Barlett2007,romero2018}. In recent years, strategies for growing the ansatz adaptively (an idea first proposed in ADAPT-VQE \cite{Grimsley_2019}) have gained popularity. They have been shown to lead to shallower circuits, higher accuracy, and improved resilience against local traps \cite{Grimsley_2023}. Our work is aimed at such adaptive algorithms. A detailed description of the workflow of ADAPT-VQE in provided in Sec.~\ref{ss:ADAPT}.

In order to solve the electronic structure problem using a quantum computer, we need a fermion-to-qubit mapping. A popular choice is the Jordan-Wigner transform \cite{JordanWigner}, given by
\begin{align}
\begin{split}
a_i^\dagger\rightarrow \frac{1}{2}\prod_{k=1}^{i-1}Z_k\cdot(X_i-iY_i),\\
\quad
a_i\rightarrow \frac{1}{2}\prod_{k=1}^{i-1}Z_k\cdot(X_i+iY_i),
\end{split}
\label{eq:jw_transform_paulis}
\end{align}
where $Z_k$, $X_i$, $Y_i$ are Pauli operators acting on the qubits labeled by the respective indices. $a_i^\dagger$ ($a_i$) is the creation (annihilation) operator for orbital $i$. This transformation can be used to map fermionic Hamiltonians to quantum-mechanical observables, as well as to transform fermionic state preparation unitaries to circuits. 

\subsection{ADAPT-VQE}
\label{ss:ADAPT}

We now introduce the relevant facets of the ADAPT-VQE algorithm, proposed in Ref~\cite{Grimsley_2019}. 

\subsubsection{Algorithm}

The idea behind ADAPT-VQE is to let the molecule under study `choose' its own state preparation circuit, by creating the ansatz in a strongly system-adapted manner. Pseudo-code for this protocol can be found in Algorithm~\ref{alg:adaptvqe}. We use $^*$ to denote optimized values.

\begin{algorithm*}
\algorithmicrequire{ \\
$\ket{\psi^{(ref)}}$, $\{\hat{A}_k\}_K$, $\mathcal{\hat{H}}$\Comment{Problem specification}\\
$\epsilon$, $L$\Comment{Hyperparameters}}\\
\algorithmicensure{\\\
$\ket{\psi^{*}}$, $x^*$, $E^*$
}
\\\dotfill\\
\everypar={\nl}
 $n\gets 0$;\\
 $x_0\gets\{\}$;\\
 $\ket{\psi^{(n)}}\gets \ket{\psi^{(ref)}}$;\\
$E_n\gets\text{measure\_energy}\left(\mathcal{\hat{H}},\ket{\psi^{(n)}}\right)$;\\
 \While{$n<L$}{{}
 $n \gets n+1$;\\
 \For{$k\gets 1...K$}{
  Measure $g_k=\bra{\psi^{(n-1)}}\left[\mathcal{\hat{H}},\hat{A}_k\right]\ket{\psi^{(n-1)}};$\label{step:meas_grads}\Comment{Measure pool gradients}\\
 }
 $i\gets k$ s.t. $g_k = \text{max}(\{|g_k|\}_K)$\Comment{Select new operator}\\
 $G \gets\Vert\{g_1,...,g_K\}\Vert_F;$\Comment{Calculate gradient norm}\\
   \eIf{$G>\epsilon$}{
   $\ket{\psi^{(n)}}\gets e^{\theta_i\hat{A_i}}\ket{\psi^{(n-1)}};$\Comment{Grow ansatz}\\
   $x_n\gets\{x_{n-1},0\}$;\Comment{Grow parameter vector}\label{alg:rec_param}\\
   $E_n, x_n\gets\text{VQE}(\mathcal{\hat{H}},\ket{\psi^{(n)}},x_n)$;\Comment{Minimize energy}\label{step:VQE_sub}\\
   }{
   Return $\ket{\psi^{(n-1)}}$, $x_{n-1}$, $E_{n-1}$;\Comment{Successful convergence}
   }
 }
 Return $\ket{\psi^{(n)}(x)}$, $x_n$, $E_n$;\Comment{Convergence target unmet}
\caption{ADAPT-VQE}
\label{alg:adaptvqe}
\end{algorithm*}

Here, $\epsilon$, $L$ are user-specified hyperparameters that define termination. The algorithm stops when the norm of the pool gradients is below $\epsilon$ or when a maximum number of iterations $L$ is reached, whichever happens first.

The role of the user in the creation of the ansatz is the selection of a size $K$ \textit{operator pool} $\{\hat{A}_k\}_K$. This pool contains the generators of the unitary operators that may be added to the ansatz. Circuits for these unitaries may be created using ladders-of-CNOTs \cite{NielsenChuang} (along with Trotterization if the corresponding Pauli strings do not commute), or pool-specific protocols \cite{Yordanov2020circuits}.

The ansatz is initialized to identity: In the first iteration, the prepared state is simply the (classical) Hartree-Fock ground state. Each iteration adds an operator to the ansatz, along with the corresponding variational parameter, initialized at zero. Thus, the state preparation circuit and the parameter vector grow from iteration to iteration.

The selected pool operator is the one which leads to the derivative $\pdv{E^{(n)}}{\theta_k}$ of greatest magnitude at point $\theta_k=0$. These derivatives can be written as an expectation value using the formula
\begin{equation}
    \pdv{E^{(n)}}{\theta_k}\Bigg|_{\theta_k=0} = \bra{\psi^{(n-1)}}\left[\mathcal{\hat{H}},\hat{A}_k\right]\ket{\psi^{(n-1)}}.
    \label{eq:comm_adapt}
\end{equation}

In each iteration $n$, the VQE subroutine (step \ref{step:VQE_sub}) minimizes the energy with respect to a fixed structure ansatz $\ket{\psi^{(n)}}$ containing $n$ variational parameters. The initial point for the optimization is the previous iteration's optimized vector augmented with a zero (step \ref{alg:rec_param}), which has been shown to improve trainability and resilience against local minima as compared to random initialization \cite{Grimsley_2023}.

This parameter value recycling strategy can be used independently of the rest of the algorithm. We can bypass the dynamic circuit creation and grow it in a predefined manner, albeit still augmenting the parameter vector iteratively. This approach falls within the realm of optimization strategies rather than ansatz design, but it can be seen as an adaptive state preparation scheme with a trivial selection criterion. This has been successfully applied to tasks such as classifying hand-written digits with quantum neural networks \cite{skolik2020} and finding the maximum cut value on a graph using the quantum approximate optimization algorithm (QAOA) \cite{lee2021}.

\subsubsection{Operator Pool}

The operator pool restricts the type of ansätze ADAPT-VQE can construct, 
and thus is the most important user-defined aspect of the algorithm. Currently, the most hardware-efficient operator pools are the qubit excitation (QE) pool \cite{yordanov2020} and the qubit pool \cite{Tang2021}. 

The qubit excitation pool is comprised of two- or four-qubit operators which preserve particle number and Z spin projection ($S_z$), but do not respect the fermionic anticommutation relations. An example QE acting on four spin-orbitals $p$, $q$, $r$, $s$ is 

\begin{align}
\begin{split}
\hat{\tau} = i(
&-X_qX_pX_sY_r
-X_qX_pY_sX_r\\
&+X_qY_pX_sX_r
-X_qY_pY_sY_r\\
&+Y_qX_pX_sX_r
-Y_qX_pY_sY_r\\
&+Y_qY_pX_sY_r
+Y_qY_pY_sX_r
).
\end{split}
\label{eq:qubit_excitation}
\end{align}
Efficient circuit implementations for QE evolutions were proposed in Ref.~\cite{Yordanov2020circuits}.

Qubit pools \cite{Tang2021} are pools in which each operator consists of an individual Pauli string. They do not conserve particle number or $S_z$ in general, nor do they respect anticommutation. The corresponding evolutions are straightforwardly implemented using ladder-of-CNOTs circuits \cite{NielsenChuang}. We consider the qubit pool formed from all individual Pauli strings appearing in the QE pool.

These two pools define two subclasses of the ADAPT-VQE algorithm: the Qubit Excitation Based (QEB)-ADAPT-VQE \cite{yordanov2020} and the Qubit-ADAPT-VQE \cite{Tang2021}. We note that the former was proposed with a few possible algorithmic modifications in addition to the choice of pool. However, since such modifications are outside of the scope of this work, we take it to be the canonical ADAPT-VQE protocol (as defined in Algorithm \ref{alg:adaptvqe}) implemented with the QE pool.

\subsubsection{Measurement Costs}
\label{sss:adapt_meas_costs}

One of the main limitations of adaptive VQAs is the scaling of the measurement costs. These costs come from two components: the VQE subroutine (step \ref{step:VQE_sub}) and the measurement of the pool gradients (step \ref{step:meas_grads}). A brief discussion of these costs follows.

The molecular Hamiltonian of a system represented by $N$ spin-orbitals (qubits) will have $\mathcal{O}(N^4)$ terms, so this is the worst-case measurement cost of an energy evaluation. However, empirical evidence shows that the Pauli strings in molecular Hamiltonians may be grouped into commuting sets of linear size, resulting in an $\mathcal{O}(N^3)$ cost for each energy evaluation (which seems unlikely to be further decreased) \cite{Gokhale2020,yen2020}. The total measurement cost of one optimization is the cost of one energy evaluation multiplied by the total number of energy evaluations, which might come directly from energy measurements or indirectly from gradient measurements (measuring a length $n$ gradient vector comes at a cost of $2n$ energy evaluations; see Appendix~\ref{ap:psrs} for a review on how to measure gradients on hardware). 

As for the cost of measuring the gradients (step \ref{step:meas_grads}), it is $\mathcal{O}(N^5)$ for both the qubit and QE pools \cite{anastasiou2023}. This means that the bound on the optimization costs is higher than the bound on the gradient measurement costs if the number of energy measurements per optimization grows faster than quadratically with $N$. Typical implementations of line searches require evaluating the gradient vector at least
once (see Appendix~\ref{ap:line_search} for a description of a line search algorithm). In this case, the optimization costs will dominate if the number of optimizer iterations (line searches) grows faster than linearly with $N$. We verify this is the case in numerical simulations. 

As such, we believe that strategies to expedite the optimization---such as ours---tackle the biggest source of measurement costs in this algorithm as of now. We confirm this numerically in the results section (Table \ref{tab:costs}).

\subsection{Gradient-Based Optimization Methods}

The choice of classical optimizer is pivotal in a VQA: it impacts not only the quality of the output solution, but also the costs, since different optimizers will require different numbers of calls to the quantum computer. We refer to Refs.~\cite{Tilly_2022, Napp_2020} for an overview of the topic.

We focus on numerical optimization methods which explicitly use the  gradient vector when setting the search direction. We refer to Appendix \ref{ap:psrs} for a discussion on how to evaluate the gradient when the cost function is evaluated using a quantum computer.

Gradient-based optimization methods employ a succession of line searches, using the final point of each as the initial point for the next. $p_k$, the search direction at iteration $k$, is determined from data collected at $x_k$, the initial point. Such data includes, but is not necessarily limited to, the gradient vector. The optimizer then seeks a step size $\alpha$ which (perhaps approximately) minimizes the cost function $f$ along $p_k$. The next iterate is set as
\begin{equation}
x_{k+1}=x_k+\alpha p_k.
\label{eq:line_search}
\end{equation}
In the next sections, we discuss different strategies to choose $p_k$.

\subsubsection{Gradient Descent}

Vanilla gradient descent \cite{num_optimization} is a first-order optimization method where the search direction at iteration $k$ is opposite to the gradient at $x_k$, i.e.,
\begin{equation}
p_k^{(GD)} = -\nabla f(x_k).
\label{eq:pk_gdescent}
\end{equation}
This means that steps are taken in the direction of steepest descent. Note that there always exists an $\alpha>0$ such that this step direction produces a lower value of $f$ at the next iterate (Eq.~\eqref{eq:line_search}). Since the gradient is not suggestive of a particular value for $\alpha$, heuristics must be used.

Gradient descent usually takes significantly longer to converge than more sophisticated alternatives.

\subsubsection{Newton's Method}

Newton's method \cite{num_optimization, Flet_87, Fletcher_1982} employs a second-order modification to the gradient descent direction using $\nabla^2 f$, the Hessian of $f$ (a matrix containing all its second derivatives). Newton's direction, 
\begin{equation}
p_k^{(N)} = -\nabla^2 f^{-1}(x_k)\nabla f(x_k),
\label{eq:pk_newton}
\end{equation}
is the vector pointing at the minimum of a quadratic model of $f$ at $x_k$, as given by a second-order Taylor expansion,
\begin{equation}
    \begin{split}
    f(x_{k}+\Delta x) \approx
    &f(x_{k}) + \left(\nabla f(x_{k})\right)^T\Delta x \\
    &+\frac{1}{2}\left(\Delta x\right)^T\nabla^2 f(x_{k})\Delta x.
    \label{eq:taylor}
    \end{split}
\end{equation}
If $\nabla^2 f(x_{k})$ is positive definite, it admits an inverse and induces a convex quadratic model. Therefore, the positive definiteness of the Hessian guarantees that Eq.~\eqref{eq:taylor} has a minimum. If this condition does not hold, $p_k^{(N)}$ might be undefined or correspond to an ascent direction. This is more likely in regions farther away from a minimum, where a bowl-shaped approximation is bound to be less adequate. In such cases, a positive-definite modification of the Hessian will typically be used in its place.

Since a unit step size would lead to the minimum if $f$ were quadratic in its variables, most practical implementations set $\alpha$ to one (with possible adjustments depending on the observed decrease of the cost function).

The use of second-order derivatives enables Newton's method to enjoy a quadratic convergence rate \cite{More_1982}. However, this comes at considerable cost: the evaluation of $\mathcal{O}(n^2)$ second derivatives and the inversion of a $n\times n$ matrix, where $n$ is the optimization dimension.


\subsubsection{Quasi-Newton Methods}
\label{ss:qn_methods}

Quasi-Newton methods \cite{num_optimization, Flet_87, Fletcher_1982, Dennis_1977} collect and use second-order information without explicitly evaluating the second derivatives of $f$. In general, they converge slower than Newton's method, but faster than gradient descent. Despite never computing the Hessian matrix explicitly, they often reach superlinear convergence rates. In these methods, the search direction is given by
\begin{equation}
p_k^{(QN)} = -H_k\nabla f(x_k),
\label{eq:pk_quasinewton}
\end{equation}
where $H_k$ is an approximation to the inverse Hessian (the actual Hessian is usually denoted $B_k$ in numerical optimization literature). This approximation is built and updated along the optimization, using available information that does not directly include second derivatives.

There are many quasi-Newton methods. We will focus on BFGS, since it is considered the most efficient \cite{num_optimization} and has many desirable properties (see App.~\ref{ap:bfgs_properties}). This method was named after Broyden \cite{Broyden_1970}, Fletcher \cite{Fletcher_1970}, Goldfarb \cite{Goldfarb_1970} and Shanno \cite{Shanno_1970} who proposed it independently in 1970. We provide pseudo-code for this optimizer in Algorithm \ref{alg:bfgs}. While many variants exist, we choose to remain as close as possible to the implementation in Scipy's \cite{Scipy} numerical optimization submodule \texttt{optimize}, as it is widely used in practice.

\begin{algorithm*}
\SetAlgoLined
\algorithmicrequire{\\
$f$, $\nabla f$, $x_0$\Comment{Problem specification}\\
$M$, $\epsilon_o$ \Comment{Hyperparameters}}\\
\algorithmicensure{\\
minimizer $x^*$, $f(x^*)$}
\\\dotfill\\
\everypar={\nl}
$n\gets \text{length}(x_0)$\\
$H_0 \gets \mathit{I}_{n\times n}$\Comment{Initialize inverse Hessian}\\
$k \gets 0$\\
 \While{$k<$ M}{ 
   $p_{k} \gets -H_{k}\nabla f(x_{k})$;\label{step:bfgs_search_dir}\Comment{Update search direction}\\
  $x_{k+1}, f(x_{k+1}),\nabla f(x_{k+1}) \gets \texttt{line\_search}(p_k,f,\nabla f, x_k, f(x_k),\nabla f(x_k))$\;
  \eIf{$\Vert\nabla f(x_{k+1})\Vert_F>\epsilon_o$}{
   $s_k\gets x_{k+1}-x_k$; \label{step:s_k}\\
   $y_k\gets \nabla f(x_{k+1}) - \nabla f(x_k)$; \label{step:y_k}\\
   $H_{k+1}\gets \texttt{update\_h}(H_k, s_k, y_k)$;\Comment{Update inverse Hessian}\\
   $k\gets k +1$\;
   }{
   Return $x_{k+1}, f(x_{k+1})$;\Comment{Successful convergence}
  }
 }
Return $x_{k}, f(x_{k})$;\Comment{Convergence target unmet}
 \caption{BFGS}
\label{alg:bfgs}
\end{algorithm*}

$f$ and its gradient $\nabla f$ must be supplied as callables, such that the optimizer can evaluate them for any parameter vector. Common choices for the initial point $x_0$ are all-zero, random, and problem-specific initializations. 

The hyperparameters $M\in\mathbb{N}$, $\epsilon_o\in\mathbb R^+$ correspond respectively to the maximum number of iterations (line searches) and to a convergence threshold on the gradient norm. We use the subscript $o$ to distinguish this threshold from the ADAPT-VQE convergence threshold in Algorithm~\ref{alg:adaptvqe}. The optimization is stopped when the number of iterations reaches $M$ or when the magnitude of the gradient vector falls below $\epsilon_o$, whichever happens first. Typical values for $M$, $\epsilon_o$ are on the order of $10^2$ to $10^4$, $10^{-6}$ to $10^{-8}$ respectively.

\texttt{line\_search} is an algorithm which seeks the $\alpha$ that minimizes $f(x_{k+1})$, with $x_{k+1}$ given by Eq.~\eqref{eq:line_search}. We provide more details about this subroutine in App.~\ref{ap:line_search}.

In the first iteration of BFGS, the direction calculated in step \ref{step:bfgs_search_dir} is opposite to the gradient (just like in gradient descent algorithms). This is due to the choice of setting $H_0$ to the identity. In principle, this could be any symmetric positive definite matrix; however, in general, heuristics for how to choose it are lacking. The identity matrix is the standard, unbiased option, and it is usually assumed by numerical optimizers (this is the case in SciPy's implementation \cite{Scipy}).

As the algorithm proceeds, $H_k$ will be updated to better reflect the curvature of $f$, thus refining the search direction. These updates depend only on parameter and gradient vectors. For more details we refer to App.~\ref{ap:bfgs_update}.

\subsubsection{Convergence Rates}
\label{sss:qrates}

A brief discussion of convergence rates follows. We refer to Ref.~\cite{num_optimization} for details.

In this work, we will define convergence rates based on ratios of errors. This is referred to as `Q-convergence' (from \textit{quotient}) and is standard in numerical optimization literature. 

The iterates $\{x_k\}$ are said to converge Q-linearly to the solution $x^*$ if there is a constant $r\in (0,1)$ such that 
\begin{equation}
    \frac{\Vert x_{k+1} - x^*\Vert_F}{\Vert x_{k} - x^*\Vert_F}\leq r
    \label{eq:linear}
\end{equation} 
for all large enough $k$. They are said to converge Q-superlinearly if 
\begin{equation}
    \lim_{k\rightarrow\infty}\frac{\Vert x_{k+1} - x^*\Vert_F}{\Vert x_{k} - x^*\Vert_F}= 0\,,
    \label{eq:superlinear}
\end{equation}
and Q-quadratically if there is a constant $M\in \mathbb{R}^+$ such that
\begin{equation}
    \frac{\Vert x_{k+1} - x^*\Vert_F}{\Vert x_{k} - x^*\Vert_F^2}\leq M
    \label{eq:quadratic}
\end{equation} 
for all large enough k.

Q-linear convergence is typical of gradient descent, while Newton's method usually enjoys Q-quadratic convergence. Quasi-Newton methods are in between, often converging Q-superlinearly. More specifically, if the quasi-Newton Hessian $B_k$ and search direction $p_k$ obey
\begin{equation}
    \lim_{k\rightarrow 0}\frac{\Vert(B_k-\nabla^2f(x^*))p_k\Vert_F}{\Vert{p_k}\Vert_F} = 0\,,
    \label{eq:diff_along_pk}
\end{equation}
then there will be an index $k_0$ such that the unit step length will be accepted for all iterations $k>k_0$, and the iterates $\{x_k\}$ will converge to the solution Q-superlinearly (see Theorem 3.6 in Ref.~\cite{num_optimization}). 

The difference between the approximate and exact Hessians along the search direction going to zero as the iterations proceed (equation \eqref{eq:diff_along_pk}) is both necessary and sufficient for quasi-Newton methods to be Q-superlinearly convergent.

As the Q-convergence definition will be used at all times, we will omit the 'Q-' when referring to convergence rates in what follows. 

\section{BFGS Algorithm for ADAPT-VQE}
\label{s:bfgs_for_adapt}

Having discussed the BFGS optimizer as well as the ADAPT-VQE algorithm, we now propose an ADAPT-VQE-tailored BFGS optimizer in Algorithm \ref{alg:BFGS_ADAPT}.

\begin{algorithm*}
\SetAlgoLined
\algorithmicrequire{\\
$f$, $\nabla f$, $x_{(n-1)}^{*}$, $\nabla f(x_{(n-1)}^{*})$, $H_{n-1\times n-1}^*$\Comment{Problem specification}\\
$M$, $\epsilon_o$ \Comment{Hyperparameters}}\\
\algorithmicensure{\\
minimizer $x^*$, $f(x^*)$, $\nabla f(x^*)$, $H_{n\times n}^*$}
\\\dotfill\\
\everypar={\nl}
$x_0 \gets \{x_{(n-1)}^{*},0\}$\\
$g_0 \gets \{{\nabla f(x_{(n-1)}^{*}),\frac{\partial{f}}{\partial{\theta_n}}(x_0)}\}$\\
$H_0 \gets\begin{pmatrix}
 H_{n-1\times n-1}^* & 0\label{step:expand_hess}\\
0 & 1\\
\end{pmatrix}$\Comment{Initialize inverse Hessian}\\
$k \gets 0$\\
 \While{$k<$ M}{ 
   $p_{k} \gets -H_{k}\nabla f(x_{k})$;\Comment{Update search direction}\\
  $x_{k+1},f(x_{k+1}),\nabla f(x_{k+1}) \gets \texttt{line\_search}(p_k,f,\nabla f, x_k, f(x_k),\nabla f(x_k))$\;
   $s_k\gets x_{k+1}-x_k$\;
   $y_k\gets \nabla f(x_{k+1}) - \nabla f(x_k)$\;
   $H_{k+1}\gets \texttt{update\_h}(H_k, s_k, y_k)$;\label{alg3:update_hess}\Comment{Update inverse Hessian}\\
   $k\gets k +1$\;\If{$\Vert\nabla f(x_k)\Vert_F<\epsilon_o$}{
   Return $x_k, f(x_k), \nabla f(x_k), H_k$;\Comment{Successful convergence}\\
  }
   }
Return $x_{k}, f(x_{k}), \nabla f(x_k), H_k$;\Comment{Convergence target unmet}
 \caption{BFGS for ADAPT-VQE Iteration n}
\label{alg:BFGS_ADAPT}
\end{algorithm*}

In labeling the inverse Hessian $H$, we use a single subscript $k$ to denote the $k$th optimization iteration, and $n\times n$ to denote the $n$th ADAPT-VQE iteration (since this sets the dimension of the matrix). When either iteration label is clear from context, we omit the corresponding subscript to simplify the notation.

We note that aside from the callables $f$, $\nabla f$, all inputs to the algorithm at iteration $n$ are outputs from iteration $n-1$. The point $x_{(n-1)}^{*}$, as well as the corresponding gradient $\nabla f(x_{(n-1)}^{*})$ and inverse Hessian $H_{n-1\times n-1}^*$, pertain to the state of the optimizer at the end of iteration $n-1$. 

The novelty of Algorithm \ref{alg:BFGS_ADAPT} is in step \ref{step:expand_hess}, where we use the inverse Hessian resulting from the ($n-1$)th ADAPT-VQE optimization to initialize the inverse Hessian at the $n$th optimization. We note that because the final inverse Hessian will be used as an input for the next iteration, we update it in line \ref{alg3:update_hess}, before the convergence check. This is in contrast with Algorithm~\ref{alg:bfgs}, where the Hessian is not updated in the iteration where convergence is reached.

The motivation behind Algorithm~\ref{alg:BFGS_ADAPT} is that ADAPT-VQE uses the final point of one optimization as the initial point for the next (see Algorithm~\ref{alg:adaptvqe}). Since at the start of the new optimization we have not moved in the cost landscape, the curvature information we gathered during the previous optimization continues to approximately capture the shape of the parameter landscape, at least in regards to the parameters $\theta_0,...,\theta_{n-1}$. As we have no second-order information concerning $\theta_n$ yet, we choose to expand $H_{n-1\times n-1}^*$ with a unit diagonal and zeros elsewhere. This unbiased choice evidently preserves the positive definiteness of the matrix, thus our modified algorithm also guarantees that the quadratic model has a minimum and $p_k$ corresponds to a descent direction at all times. 

It is simple to see that positive definiteness is also preserved by the act of removing a row and the corresponding column from $H_k$. This means that we have the freedom to select from which parameters we wish to preserve second-order information. As a result, our strategy generalizes to the case where we wish to freeze a subset $\{\theta_{f_1},...,\theta_F\}$ of $F$ parameters that were active in the previous iteration. In this case, we simply remove from $H_{n-1\times n-1}^*$ the rows and columns corresponding to the indices $f_1,...,F$. This was not explicitly included in Algorithm~\ref{alg:BFGS_ADAPT} for the sake of conciseness.

Equipped with our modified BFGS algorithm, ADAPT-VQE has the flavor of a continuous optimization of growing dimension. The search space is expanded when the ansatz gradient norm falls below $\epsilon_o$, because we do not expect a significant energy decrease along the directions in the current cost landscape. The algorithm terminates when the pool gradient norm is below $\epsilon$, because we do not expect a significant energy decrease along the directions in which we can expand this space.

The selection criterion of ADAPT-VQE dictates that the direction in which to expand the parameter space leads to the steepest cost landscape. This is reminiscent of first-order optimization methods, with the difference being that here we are screening possible candidates, and so the choice of direction is made with respect to parameters not currently in the parameter vector.

For the sake of conciseness, we will refer to our approach as \textit{recycling the Hessian}. Note that the inverse Hessian is unique and fully determines the actual Hessian, such that by recycling one we implicitly recycle the other.

\section{Results}
\label{s:results}

In this section, we present numerical simulation results for different systems at various bond lengths. While we focus on linear H$_6$ (12 qubits) as a toy model for a strongly correlated system, we additionally consider LiH (12 qubits) and BeH$_2$ (14 qubits) as real molecules. In all cases, we use the STO-3G basis set and no frozen-core approximations. The pools we use are those defined in Sec.~\ref{ss:ADAPT}: the qubit excitation pool \cite{yordanov2020} and the qubit pool \cite{Tang2021}. We set the ADAPT-VQE convergence threshold $\epsilon$ to $10^{-6}$ and $10^{-5}$, respectively. The threshold for the qubit pool is set to a higher value because this pool is larger.

The code used for the numerical simulations has been made publicly available on GitHub \cite{ceo_repo}. OpenFermion \cite{openfermion} was used for manipulating fermionic operators, and the corresponding plugin with PySCF \cite{pyscf} for the underlying electronic structure calculations. All expectation values were calculated via matrix algebra. To calculate distances between matrices, we use the Frobenius norm, defined for an $m\times n$ matrix $A$ as $\Vert A \Vert_F = \sqrt{\text{Tr}(AA^\dagger)}=\sqrt{\sum_{i=1}^m\sum_{j=1}^n|a_{ij}|^2}$. The hyperparameters for BFGS were chosen as $\epsilon_o=10^{-6}$ and $M=10000$ (see Sec.~\ref{ss:qn_methods}). 

We now briefly describe how this section is organized. We begin by analyzing the impact of recycling the Hessian on the measurement costs of ADAPT-VQE for multiple molecules, interatomic distances, and pools in Sec.~\ref{ss:measurement_costs}. In Sec.~\ref{ss:dist_exact}, we seek to understand when our strategy works the best by analyzing the distance between the approximate and exact (inverse) Hessians, as well as the evolution of the latter. Finally, in Sec.~\ref{ss:opt_evol}, we dive into specific optimizations to better understand the faster convergence of our method.

\subsection{Measurement Costs}
\label{ss:measurement_costs}

In this subsection, we assess the impact of recycling the Hessian on the measurement costs of the optimization process. We consider the key cost to be the number of function evaluations. Evaluating the energy corresponds to one function evaluation, while evaluating an element of the gradient vector corresponds to two (see Appendix \ref{ap:psrs}). The circuits for these two cases differ by a negligible number of gates, and the observable is always the Hamiltonian \cite{Kottmann_2021}. This means that each function evaluation requires measuring the same set of Pauli strings and implies a similar number of shots, up to differences in the variance of the expectation value of the energy in the state. Note that we can place a state-independent upper bound on the number of shots required for a given error \cite{Wecker2015}. Based on these arguments, we expect the total number of function evaluations to be a good figure for assessing costs.

\begin{figure*}[htbp]
    \includegraphics[width=0.97\textwidth]{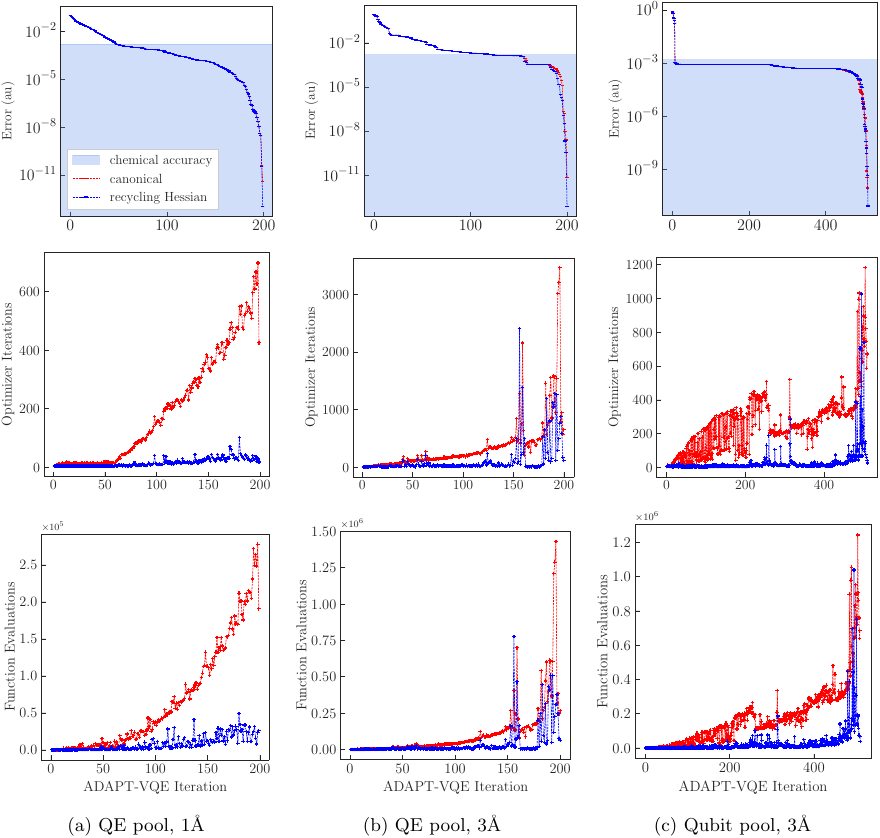}
    \caption{Comparison of the error (top row), total number of line searches (middle row), and total number of function evaluations (bottom row) per iteration of the ADAPT-VQE algorithm with and without Hessian recycling for H$_6$. We consider the QE pool for (a) equilibrium and (b) stretched geometries, as well as (c) the qubit pool for a stretched geometry. The red curve is often not visible in the error plots due to it lying below the blue one. The region shaded blue is the region of chemical accuracy, defined as an error below 1 kcal/mol.}
    \label{fig:h6_results}
\end{figure*}

Figure \ref{fig:h6_results} shows the results for the H$_6$ molecule at various bond distances, using the QE and qubit pools. The first thing to note is that the error curves overlap for nearly all iterations, except for regimes of very high accuracy where recycling the Hessian results in a lower absolute error. We do not expect this to be a benefit of our method, as this accuracy range is unlikely to be relevant in practice. Simulation data shows that up to such high accuracy regimes, the final ans\"atze are identical regardless of whether we recycle the Hessian. 

Despite the matched accuracy, the number of optimizer iterations is remarkably different between the two methods. When the initial estimate for the inverse Hessian is the identity, the number of BFGS iterations clearly increases faster than linearly with the ADAPT-VQE iteration number; in contrast, when we recycle the Hessian, the number of line searches is roughly constant across a large number of ADAPT-VQE iterations, despite the considerable increase in the dimension of the parameter space. Since the second-order information is not reconstructed from scratch in each optimization, accounting for parameter correlations does not necessarily imply a superlinear number of optimizer iterations. In fact, the search direction for the very first line search at ADAPT-VQE iteration $n$ is already equipped with information concerning correlations between $n-1$ parameters, represented by $(n-1)^2$ second derivatives. The only missing information concerns a linear number of second derivatives, describing the correlation between each parameter and the last.

Interestingly, the impact of recycling the Hessian for H$_6$ at 3\AA{} is more significant on qubit-ADAPT-VQE (Fig.~\ref{fig:h6_results}(c)) than on QEB-ADAPT-VQE (Fig.~\ref{fig:h6_results}(b)). With the qubit pool, the Hessian recycling protocol decreases the total number of function evaluations by 84\% --- 20\% more than the decrease with the QE pool. This suggests that the impact of the strategy is not only system- but also pool-dependent. 

In general, for a fixed dimension, we expect a higher number of line searches to imply a higher number of function evaluations (see Appendix \ref{ap:line_search}). This is confirmed by the bottom panels of Fig. \ref{fig:h6_results}, which show that the Hessian recycling strategy succeeds in producing a significant decrease in measurement costs. 

In Appendix~\ref{ap:other_mols}, we include additional plots for the LiH and BeH$_2$ molecules. In all cases, recycling the Hessian results in relevant savings in measurement costs. The savings are particularly notable for the H$_6$ molecule (especially at the equilibrium geometry). Among the three, this is the most difficult to simulate, requiring the most iterations and measurements. In all cases, the difference in costs seems to increase as the iterations advance. Thus, we can expect our method to become even more beneficial for larger systems.

Considering the same molecules at larger bond distances allows us to study the impact of our protocol as systems become more strongly correlated. While the savings in the number of line searches and measurements are maintained or even increased as we stretch LiH and BeH$_2$, they become less significant for H$_6$ at stretched geometries. Among the test cases we considered, this was the one where our strategy performed the worst. We investigate the causes in Sec.~\ref{ss:dist_exact}. Despite this, recycling the Hessian still results in a reduction of the total number of function evaluations by $64\%$ across the whole execution.

We finish this section with quantitative examples of the measurement cost reduction achieved by our method. Table \ref{tab:costs} includes the costs of the VQE step with and without Hessian recycling, as well as of the gradient measurement step, through complete executions of QEB-ADAPT-VQE for various systems. Note that the latter step concerns the measurement of the gradients of operators generated by pool elements, not of operators in the ansatz (which are included in the costs for VQE). We do not consider any grouping strategies for the Hamiltonian, as they are beyond the scope of this work.

\begin{table*}[htbp]
  \centering
\begin{center}
\begin{tabular}{|c|c|c|c|c|c|c|c|c|}
\cline{4-9}
\multicolumn{3}{c|}{} & \multicolumn{6}{c|}{Molecule}\\
\cline{4-9}
\multicolumn{3}{c|}{} & \multicolumn{2}{c|}{LiH} & \multicolumn{2}{c|}{H$_6$} & \multicolumn{2}{c|}{BeH$_2$} \\
\cline{4-9}
  \multicolumn{3}{c|}{} & 1.5\AA{} & 3\AA{} & 1\AA{} & 3\AA{} & 1.3\AA{} & 3\AA{}  \\ 
 \hline
  \multirow{4}{*}{\rotatebox[origin=c]{90}{Step}} &\multicolumn{2}{c|}{Gradient Measurement} & $5.2\times 10^3$ & $5.4\times 10^3$ & $1.9\times 10^4$ & $1.9\times 10^4$ & $1.2\times 10^4$ & $1.3\times 10^4$ \\  
\cline{2-9}
 &\multirow{3}{*}{\rotatebox[origin=c]{90}{VQE}}&Canonical & $2.5\times 10^5$ & $2.4\times 10^5$ & $1.3\times 10^7$ & $2.2\times 10^7$ & $4.2\times 10^6$ & $3.6\times 10^6$  \\
\cline{3-9}
 &&Recycling Hessian & $6.0\times 10^4$ & $3.2\times 10^4$ & $1.7\times 10^6$ & $7.9\times 10^6$ & $9.4\times 10^5$ & $5.6\times 10^5$ \\
 &&(reduced to) & (24\%) & (13\%) & (13\%) & (36\%) & (22\%) & (16\%) \\
\hline
\end{tabular}
\end{center}
  \caption{Total measurement costs incurred by the QEB-ADAPT-VQE algorithm in the gradient measurement and VQE subroutines (steps \ref{step:meas_grads} and \ref{step:VQE_sub} in Algorithm \ref{alg:adaptvqe}), for the studied test cases. The costs are given as multipliers for the cost of a naive energy evaluation, so that unit cost implies $\mathcal{O}(N^4)$ measurements. For the gradient measurement step, we consider the worst-case cost under the leading measurement strategy for the QE pool ($8N$ per iteration, where $N$ is the number of spin-orbitals/qubits) \cite{anastasiou2023}.}
  \label{tab:costs}
\end{table*}

The numerical data shows that the measurement cost of the VQE subroutine is reduced by roughly one order of magnitude when our Hessian recycling strategy is employed. We additionally verify that, as predicted by our analysis in Sec.~\ref{sss:adapt_meas_costs}, this subroutine is the bottleneck of the algorithm as far as measurement costs are concerned. In fact, the associated cost is always at least one order of magnitude higher than the cost of the gradient measurements, even when the Hessian recycling strategy is employed. As such, we confirm that our proposal addresses the most significant source of measurement costs of ADAPT-VQE, decreasing its \textit{total} measurement costs by an order of magnitude (for the studied molecules---as discussed, we expect this decrease to become more significant for larger systems). For the majority of the molecules we considered, the reduction in the measurement costs of the optimization was in the 70-90\% range.

\subsection{Distance to the Exact Hessian}
\label{ss:dist_exact}

In this subsection, we analyze the distance between the approximate and exact inverse Hessians, $H^{(\text{opt})}$ and $H^{(\text{exact})}$, with and without Hessian recycling. 

\begin{figure*}
    \includegraphics[width=\textwidth]{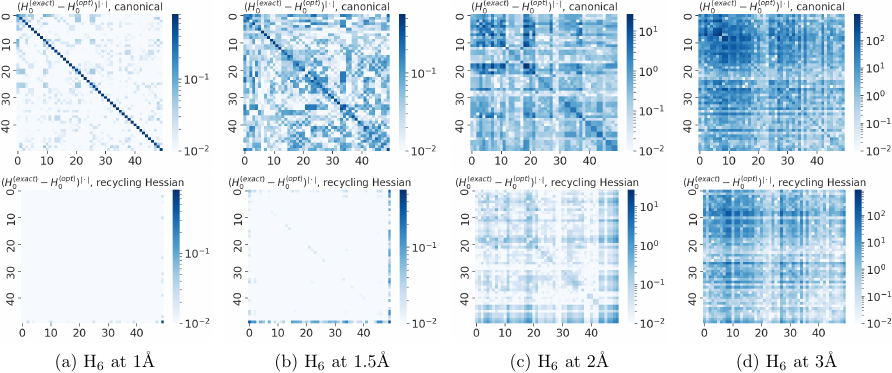}
    \caption{Heatmaps showing the difference between the initial approximate inverse Hessian in the optimization process and the initial exact inverse Hessian, with (bottom row) and without (top row) Hessian recycling, for the 50th iteration of QEB-ADAPT-VQE. The plots show the element-wise difference between these two matrices. Due to it being strongly correlated and the hardest to simulate, here we consider the test molecule H$_6$ at different interatomic distances.}
    \label{fig:heatmaps}
\end{figure*}

We begin with the heatmaps of Fig. \ref{fig:heatmaps}, a visual representation of this distance at the beginning of the 50th optimization of ADAPT-VQE with the QE pool. We plot $( H_0^{(\text{exact})}-H_0^{(\text{opt})})^{|\cdot|}$, the element-wise distance between $H_0^{(\text{opt})}$ and $H_0^{(\text{exact})}$. We consider  H$_6$ at four different bond distances, and we use the same color map for each bond distance. It should be noted that there is nothing unique about the 50th optimization. We chose this number due to it being high enough for the optimization process to be interesting, but not so high that the calculations become intractable or the heatmaps illegible. In Appendix~\ref{ap:extra_hms} we show that the behavior generalizes by presenting equivalent heatmaps for other iterations.

We note that due to how we initialize $H_k$ in Algorithm \ref{alg:BFGS_ADAPT}, the last row/column of the heatmaps is the same regardless of whether we recycle the Hessian or not. However, as expected, the inner second derivatives are better approximated by the previous iterations' derivatives than by the elements of the identity matrix. The difference is more significant for geometries near equilibrium.

Another interesting geometry-related trend can be observed in the upper heatmaps: for less correlated geometries (shorter bond distances), the approximation is the poorest for the diagonal elements \footnote{One could hypothesize that this is a matter of scaling. However, the errors in the diagonal remain dominant even if we multiply the identity matrix by a scalar factor tuned to minimize the distance to the true values.}. As we increase the bond distance, there is a shift in behavior, as the magnitude of some off-diagonal elements in the difference matrix starts rivaling the magnitude of the diagonal elements. At 3\AA{}, the diagonal entries are no longer dominant. This is clearly symptomatic of a more complicated optimization: While for shorter bond distances the parameters can nearly be treated as uncorrelated, correlations between parameters play a bigger role for stretched geometries. Relatedly, the magnitude of the elements of the difference matrix changes from well below unity up to several hundred as we increase the bond distance.

\begin{figure}[htbp]
    
    \includegraphics[width=0.64\columnwidth]{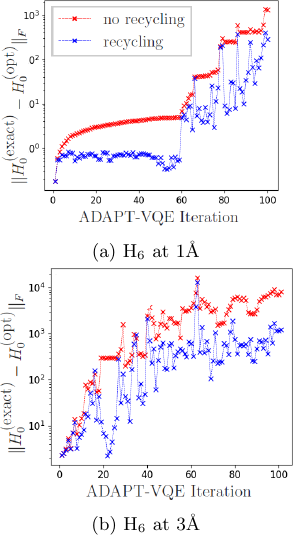}
    \label{fig:H6_qe_frobenius_r_id}
     \caption{Frobenius distance between the initial approximate inverse Hessian in the optimization process $H_0^{(\text{opt})}$ and the initial exact inverse Hessian $H_0^{(\text{exact})}$, with and without Hessian recycling, for the first hundred QEB-ADAPT-VQE iterations. We consider two cases: one for which the Hessian recycling works particularly well (H$_6$ at 1\AA{}) and one for which it does not (H$_6$ at 3\AA{}).}
\end{figure}

While the heatmaps may help us gain intuition, the information they provide is limited, as they only concern one iteration. In Fig. \ref{fig:H6_qe_frobenius_r_id}, we plot the Frobenius distance between these same matrices for the first 100 iterations of QEB-ADAPT-VQE. We focus on the two extreme bond distances: 1\AA{} and 3\AA{}. As expected, the initial distance to the exact inverse Hessian is greater for the stretched bond distance, both when we recycle the Hessian and when we do not. Remarkably, the impact of recycling the Hessian on the initial distance is a similar multiplicative factor (close to 0.1) for both bond distances, despite the impact on costs being significantly larger for the 1\AA{} geometry (as we saw in Figs.~\ref{fig:h6_results}(a), \ref{fig:h6_results}(b)).

\begin{figure}[htbp]
    \includegraphics[width=0.64\columnwidth]{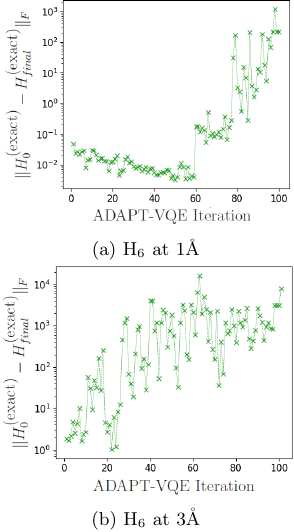}
     \caption{Frobenius distance between the exact initial $H_0^{(\text{exact})}$ and final  $H_{\text{final}}^{(\text{exact})}$ inverse Hessians, for the first hundred QEB-ADAPT-VQE iterations. The systems are the same as those from Fig. \ref{fig:H6_qe_frobenius_r_id}.}
    \label{fig:H6_qe_frobenius_evolution}
\end{figure}

We hypothesize that in large part, the difference in performance is not due to the initial matrix being a poor approximation, but rather due to the optimization process being more difficult. If the initial point is farther away from a minimum, and the curvature around these two points is vastly different, the optimizer will require more iterations to move along the cost landscape, update the inverse Hessian, and reach the minimum. For such lengthier optimizations, the cost for the optimizer to move across the landscape is more likely to surpass the cost of approximating the second derivatives around the initial point. To test this hypothesis, we plot the distance between the exact initial and exact final inverse Hessians in Fig. \ref{fig:H6_qe_frobenius_evolution}.

The plot confirms that there is a significant increase in the distance between the initial and final inverse Hessians when the bond distance is increased. For H$_6$ at 1\AA{}, the distance varies between $10^{-2}$ and $10^{3}$ and only surpasses unity after 80 iterations. In contrast, at 3\AA{}, the distance varies between $10^{0}$ and $10^{4}$ and is above unity (and closer to the maximum value) throughout all the ADAPT-VQE iterations.

\subsection{Evolution of the Optimization}
\label{ss:opt_evol}

In the previous subsections, we focused on data from the beginning or end of each optimization. This allowed us to characterize the behavior of our protocol throughout complete runs of the ADAPT-VQE algorithm, and analyze how the dimension of the optimization and the molecule under study impact the costs.

However, as the protocol we propose is in fact a numerical optimization method, it is interesting to zoom in on the optimization process and investigate the impact of recycling the Hessian in quantitative aspects of it. Thus, in this section we delve into a particular optimization. We consider the 75th iteration of QEB-ADAPT-VQE for H$_6$ at equilibrium (1\AA{}) and stretched (3\AA{}) bond distances (the same systems we analyzed in the previous section). The corresponding optimization has 75 parameters. Once again, we note that there is nothing unique to the 75th iteration; we chose it because it is complex enough to be relevant, but not so much that calculating the relevant data becomes a computational challenge. In Appendix~\ref{ap:it_50} we consider a different iteration to show that the results generalize.

\begin{figure*}[htbp]
     
    \includegraphics[width=\textwidth]{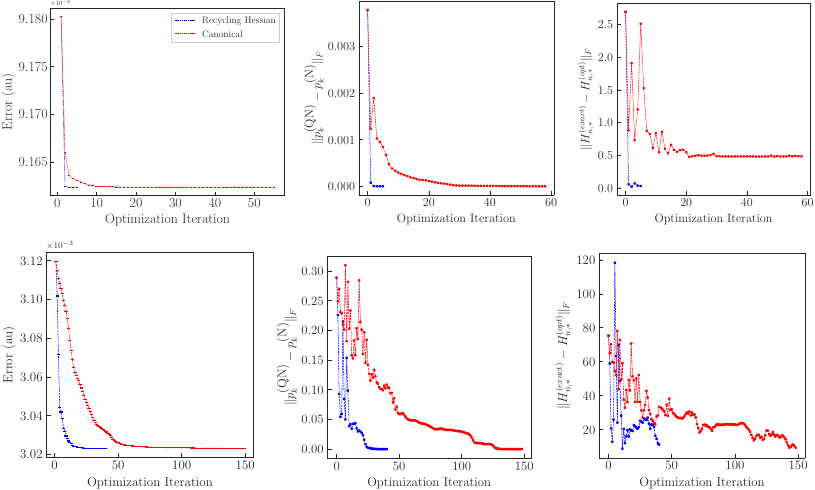}
     \caption{Evolution of the 75th QEB-ADAPT-VQE optimization with and without Hessian recycling, for the same two systems as in Fig. \ref{fig:H6_qe_frobenius_r_id}: H$_6$ at 1\AA{}  (top row) and 3\AA{}  (bottom row). The panels show (left to right) the error; the distance between the quasi-Newton $p_k^\textnormal{(QN)}$ and Newton $p_k^\textnormal{(N)}$ directions; and the distance between the last row of the approximate inverse Hessian $H^{(\text{opt})}$ and the exact inverse Hessian $H^{(\text{exact})}$.}
    \label{fig:H6_qe_it75}
\end{figure*}

In the first column of Fig.~\ref{fig:H6_qe_it75} we observe that, as expected, the optimization requires significantly fewer iterations to converge for the equilibrium geometry. In line with previous results, recycling the Hessian results in similar final error for a lower number of iterations.

In the second column, we can see that the search direction (Eq.~\eqref{eq:pk_quasinewton}) aligns much more quickly with the Newton direction (Eq.~\eqref{eq:pk_newton}) when the Hessian is recycled. We note that the initial distance to Newton's vector (which we plot as the value for iteration 0) is identical whether the Hessian is recycled or not: the $n-1$ interior parameters were previously optimized and thus have nearly zero gradients, and step \ref{step:expand_hess} in Algorithm \ref{alg:BFGS_ADAPT} is agnostic to any second-order information concerning the $n$th parameter. Despite being so similar initially, once the parameters start changing, the second-order information contained within the recycled Hessian becomes relevant and allows the direction to more quickly align with Newton's. This is remarkable: Unlike Newton's method, our optimization method never explicitly measures the Hessian, yet it aligns with the Newton search direction in a fraction of the iterations needed for the canonical BFGS implementation.

\begin{figure*}[htbp]
    \includegraphics[width=\textwidth]{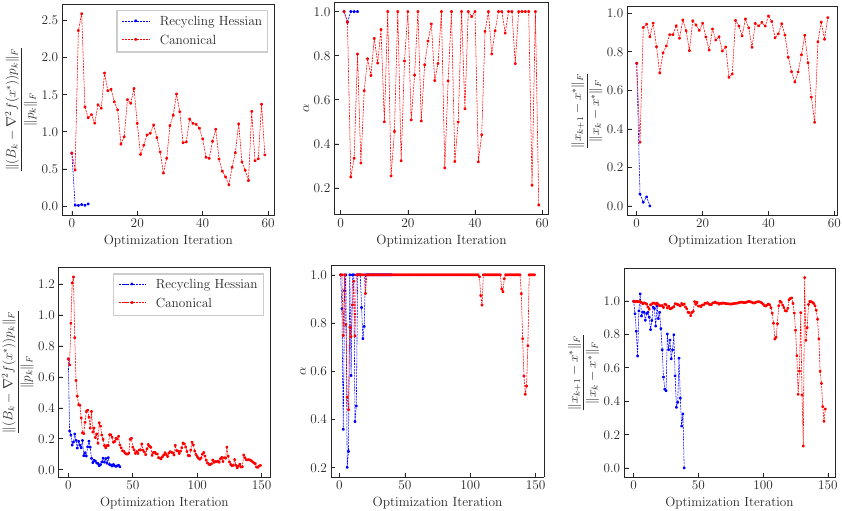}
    \caption{Relevant quantities for the study of the convergence rate of the 75th QEB-ADAPT-VQE optimization with and without Hessian recycling, for the same two systems as in Fig. \ref{fig:H6_qe_frobenius_r_id}: H$_6$ at 1\AA{}  (top row) and 3\AA{}  (bottom row). (a) The distance between the approximate and the exact Hessian along the search direction, (b) the step size $\alpha$ at the end of the line search, and (c) the ratio between successive errors are shown.}
    \label{fig:H6_qe_it75_convergence}
\end{figure*}

The third column of Fig.~\ref{fig:H6_qe_it75} shows the distance between the last row of the approximate and exact inverse Hessians. We denote the last row by $H_{n,*}$; recall that the matrix is symmetric, so that we have $H_{n,*}=H_{*,n}$. This is the vector of second-order derivatives involving the new parameter. For similar reasons to those explained in the discussion of Fig.~\ref{fig:H6_qe_it75}(b), the initial distance, plotted at iteration 0, is identical regardless of whether the Hessian is recycled or not. Since the second-order derivatives being recycled only involve the parameters $\theta_1,...,\theta_{n-1}$, naively, there seems to be no reason to expect that recycling the Hessian will lead to faster convergence of the last column/row (corresponding to $\theta_n$). Surprisingly, this does happen; it seems that the initial curvature information concerning the interior parameters allows the optimizer to focus on exploring the new search direction, and enables a faster build up of information regarding correlations between the new parameter and old ones. Even more surprisingly, the distance is still higher for the canonical BFGS when convergence is reached, despite the larger number of iterations. This is particularly visible for the 1\AA{} geometry.

Finally, we focus on the convergence rate of the optimizations. The relevant quantities (see Sec.~\ref{sss:qrates}) are plotted in Fig. \ref{fig:H6_qe_it75_convergence}. In the first column, we see that the difference between the approximate and exact Hessians along the search direction goes to zero in both optimizations when we recycle the Hessian, but this does not happen when we do not. Further, the step size (second column) saturates to unity after roughly half of the iterations when we recycle the Hessian, but oscillates instead of stabilizing when we do not. Together these results indicate that recycling the Hessian results in a superlinear convergence (Eq.~\eqref{eq:superlinear}) that would otherwise not be achieved.

Finally, the third column of Fig.~\ref{fig:H6_qe_it75_convergence} confirms this. BFGS with Hessian recycling enjoys superlinear convergence, while canonical BFGS converges linearly with a convergence constant $r$ close to 1---the worst possible scenario (see Eq.~\eqref{eq:linear}). Such a convergence rate is expected of gradient descent in ill-conditioned problems, and constitutes an underwhelming performance for a quasi-Newton optimizer from which we hope to achieve superlinear convergence.

\section{Conclusion}
\label{s:conclusion}

In this work, we proposed to tailor BFGS, one of the most popular optimizers for variational quantum algorithms, to ADAPT-VQE, one of the most popular such algorithms. In a typical implementation, each iteration of ADAPT-VQE performs an optimization where the state of the optimizer is initialized as if we possessed no knowledge about the curvature of the cost landscape, despite the fact that some knowledge was collected along the previous optimization. We develop a variant of the BFGS method that allows second-order information to naturally flow from one iteration to the next. By recycling the inverse Hessian matrix maintained by the optimizer, this protocol converges superlinearly even in situations where the canonical BFGS implementation does not, and achieves significant savings in the number of function evaluations required in each iteration of ADAPT-VQE. This specifically addresses the costs of the optimization process, which is the biggest source of measurement costs in such algorithms. In addition to decreasing the number of calls to the quantum computer in hardware implementations of adaptive VQAs, our strategy reduces the runtime of classically simulating them, thereby improving our ability to design and test these algorithms.

Since the impact of our Hessian recycling strategy seems to increase with the size of the system and the dimension of the optimization, we expect it to be even more beneficial for molecules for which performing classical simulations is infeasible.

While we focused on ADAPT-VQE for testing our proposed optimizer, our algorithm easily generalizes to other iterative state preparation protocols such as ADAPT-QAOA \cite{Zhu2020} or layerwise learning \cite{skolik2020}. Freezing parameters or layering \cite{Anastasiou2022,long_2023} are also compatible with our strategy, requiring nothing more than a simple manipulation of the initial inverse Hessian (removing or adding subsets of rows and columns).

The noise-resilience of the proposed optimization algorithm is left as an open question. Adding realistic noise to the simulations implies an overhead in computational costs which leads to prohibitive simulation times for interesting molecules. Previous approaches to the noisy simulation of ADAPT-VQEs have averted this issue by growing the ansatz noiselessly and assessing the impact of noise in the final circuit exclusively \cite{Dalton2022,long_2023}. While this may provide insight regarding the near-term viability of such algorithms, it is evidently not applicable to the optimization process. Developing strategies for noisy simulation that address this issue is left for future work. We note that while there is a general expectation that optimizers designed to be robust in the presence of noise (such as SPSA \cite{SPSA})
will be better suited for experiments in quantum hardware, BFGS was shown in Ref.~\cite{Pellow_Jarman_2021} to be among the best algorithms for noisy optimizations, with a performance comparable to SPSA. 

\section*{Acknowledgments}

We thank Raffaele Santagati and Matthias Degroote for their support and encouragement at the start of this project, and Ernesto Galvão for helpful discussions. This work is in part financed by National Funds through the Portuguese funding agency, FCT - Fundação para a Ciência e a Tecnologia, within project LA/P/0063/2020 (\href{https://doi.org/10.54499/LA/P/0063/2020}{https://doi.org/10.54499/LA/P/0063/2020}). MR acknowledges support from INESC TEC and FCT under PhD research scolarships 9575/BI-M-ED\_B2/2022 and 2022.12333.BD respectively. NJM acknowledges support from the US Department of Energy (Grant No. 	DE-SC0024619). EB acknowledges support from the US Department of
Energy (Grant No. DE-SC0022389). SEE acknowledges support by Wellcome Leap as part of the Quantum for Bio Program.

\appendix

\section{Evaluating Gradients on Quantum Hardware}
\label{ap:psrs}

The gradient function $\nabla f$ which must be supplied to gradient-based optimization methods merits a discussion, since it is not evident that this information is available when the cost function is evaluated on a quantum computer.

We note that some implementations of BFGS (including  SciPy's \cite{Scipy}) allow the user to bypass the construction of a gradient function by implementing it internally via finite difference (FD) methods. These numerical methods approximate derivatives from evaluations of the function at points which differ by small shifts, and can be applied to any function in a black-box fashion \cite{Grossmann2007}. However, when the cost function is the expectation value of a (generic) quantum mechanical observable, it must be obtained by averaging over a finite number of samples, and thus it is inevitably noisy. In addition to this we have hardware limitations inherent to NISQ computers, such as miscalibrated rotation gates and other sources of noise. All of this hampers the task of gauging minute shifts in the function value, and the fact that the finite difference quotient must have a small denominator (because its magnitude is related to the magnitude of the error of the approximation) aggravates these problems. Expectably, BFGS with FD methods has been found to perform poorly in the presence of noise \cite{Lavrijsen_2020}.

A realistic alternative to FD methods are parameter-shift rules (PSRs), proposed in Ref.~\cite{Mitarai2018} and extended in Ref.~\cite{Schuld_2019}. PSRs use a clever manipulation of analytical expressions to express the derivatives as linear combinations of measurable expectation values. Unlike FD formulas, which are generic numerical approximations, PSRs are analytical derivatives with a circuit-specific structure. When the qubit pool is used, the generators of the ansatz elements consist of a single Pauli string, whose gradient can be obtained from two energy measurements using the simplest PSRs. When the QE pool is used, more sophisticated techniques are required because each of the generators has three eigenvalues. In this case, the gradients can be measured using the fermionic PSRs proposed in Ref.~\cite{Kottmann_2021}. For real wave functions (as we are concerned with), this will similarly imply the measurement of two expectation values, with the corresponding circuits having a negligible increase in circuit depth with respect to the energy measurement circuits.

\section{BFGS Optimizer}
\label{ap:BFGS}

\subsection{Line Search Subroutine}
\label{ap:line_search}
Algorithm \ref{alg:bfgs} in the main text requires a \texttt{line\_search} subroutine to choose the next iterate. The task of this algorithm is to find the step size $\alpha$ which minimizes the function $f$ along the search direction $p_k$,
\begin{equation}
\min_{\alpha>0} f(x_k+\alpha p_k).
\label{eq:line_search_prob}
\end{equation}
We now briefly describe this subroutine.

Because finding the minimum with high accuracy might be unnecessarily costly, this minimization is often approximate and terminates when some reasonable conditions are met. A common choice are the Wolfe conditions,
\begin{equation}
f(x_k + \alpha_k p_k) \le f(x_k) + c_1\alpha_k\nabla f(x_k)^Tp_k,
\label{eq:wolfe1}
\end{equation}
\begin{equation}
\nabla f(x_k + \alpha_k p_k)^Tp_k \ge c_2 \nabla f(x_k)^T p_k,
\label{eq:wolfe2}
\end{equation}
where $0<c_1<c_2<1$. Example values for quasi-Newton methods are $c_1=10^{-4}$, $c_2=0.9$ \cite{num_optimization}. Note that if $p_k$ is a descent direction, as should be, then $\nabla f(x_k)^Tp_k$ is negative. 

Condition \eqref{eq:wolfe1}, known as the \textit{sufficient decrease} or \textit{Armijo} condition, asserts that the decrease in the value of $f$ is lower (and so higher in magnitude) than $f$'s instantaneous rate of change along $p_k$ at the initial point, weighed by $c_1$. This condition is evidently satisfied for small enough $\alpha_k$, even if the decrease in the function value is negligible. 

Condition \eqref{eq:wolfe2}, known as the \textit{curvature} condition, asserts that the derivative (along $p_k$) of the function at the new point is higher (and so lower in magnitude) than the one at the initial point, weighed by $c_2$. This bound is placed because the higher the magnitude of the rate of change of $f$, the higher the decrease in $f$ we expect from further refining the step $\alpha_k$.

Another important choice in a line search algorithm is how to initialize and vary $\alpha$. One option is to start with a guess and increase it until we find a point which either satisfies the desired conditions or brackets points which do. In the latter case, we can backtrack, decreasing $\alpha$ until a valid point is found.

The cost of this subroutine depends on how many attempts it takes us to find an $\alpha$ that satisfies Wolfe's conditions (\eqref{eq:wolfe1}, \eqref{eq:wolfe2}), which depends on the shape of the cost function and the proximity to a minimum. Each iteration of the line search requires measuring the cost function and the gradient vector, which imply respectively $1$ and $2n$ energy evaluations for an $n$-dimensional ADAPT-VQE optimization. 

\subsection{BFGS Update Rule}
\label{ap:bfgs_update}

Another subroutine required by algorithm \ref{alg:bfgs} is \texttt{update\_h}, which updates the inverse Hessian $H_k$ at the end of a line search. This update depends on the vectors $s_k$ and $y_k$, obtained from the difference between the coefficient and gradient vectors, respectively, at the $k$th and ($k-1$)th iterations (see steps \ref{step:s_k} and \ref{step:y_k} of Algorithm \ref{alg:bfgs}).

The BFGS update rule is obtained by enforcing three conditions:

\begin{enumerate}
    \item The gradient of the new quadratic model for $f$ (defined by a second-order Trotter expansion) matches the true gradient at the last two iterates.
    \item $H_{k+1}$ is symmetric and positive definite.
    \item Among all matrices satisfying the above, $H_{k+1}$ is the one which minimizes the distance to $H_k$, with respect to some norm. 
\end{enumerate}

These three conditions give rise to the BFGS update rule
\begin{equation}
H_{k+1}=(I-\rho_ks_ky_k^T)H_k(I-\rho_ky_ks_k^T) + \rho_ks_ks_k^T,
\label{eq:bfgs_update}
\end{equation}

where $\rho_k=\frac{1}{y_k^Ts_k}$. Conveniently, the BFGS method deals directly with the \textit{inverse} Hessian, required to decide the next search direction via Eq.~\ref{eq:pk_quasinewton} of the main text. Other quasi-Newton formulas, such as DFP (proposed by Davidon in 1959 \cite{Davidon_1959} and further analyzed by Fletcher and Powell \cite{Fletcher_1963}), handle the inversion posteriorly.

We note that the approximate inverse Hessian $H$ (and the induced Hessian $H^{-1}$) will be positive definite even if the same is not true for the real Hessian (and its inverse).

\subsection{Properties of BFGS}
\label{ap:bfgs_properties}

BFGS is a robust optimizer with good convergence \cite{powell_1975} and self-correcting properties \cite{Nocedal_1992}. Despite never directly evaluating second-order derivatives, it enjoys a super-linear convergence rate and performs well in practice, reaching a minimum sufficiently fast for most practical purposes \cite{Dennis_1977,num_optimization}. Because of such features, this optimizer has become a popular choice for VQAs and is often used in practical implementations \cite{guerreschi_2017,Lotshaw_2021,Grimsley_2019,Yordanov_2021,mbeng_2019}.

It has been proved that when implemented with Wolfe line searches (see App.\ref{ap:line_search}), this optimizer is globally convergent for convex functions, i.e. iterates will converge to a minimum regardless of the initial point \cite{powell_1975}. In the context of numerical optimization, \textit{global} is typically used to stress the independence of the convergence on the initial point \cite{Sriperumbudur_2012}. Accordingly, local convergence properties are those that only hold when the initial point is close enough to a minimum. We note that in this context, the words `local' and `global' do not refer to the type of minimum. Globally convergent methods may converge to a local minimum and vice-versa.

\section{Results for Other Molecules}
\label{ap:other_mols}

\begin{figure*}[htbp]
    
    \includegraphics[width=\textwidth]{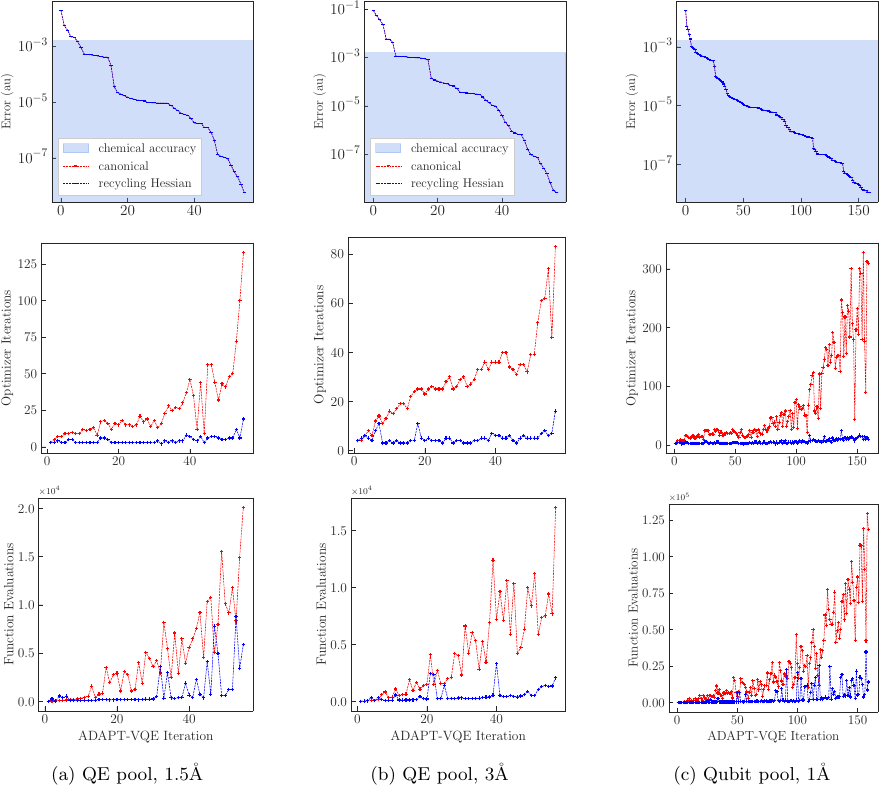}
    \caption{Equivalent plots to Fig.~\ref{fig:h6_results} for LiH.}
    \label{fig:lih_results}
\end{figure*}

\begin{figure*}[htbp]
    
    \includegraphics[width=\textwidth]{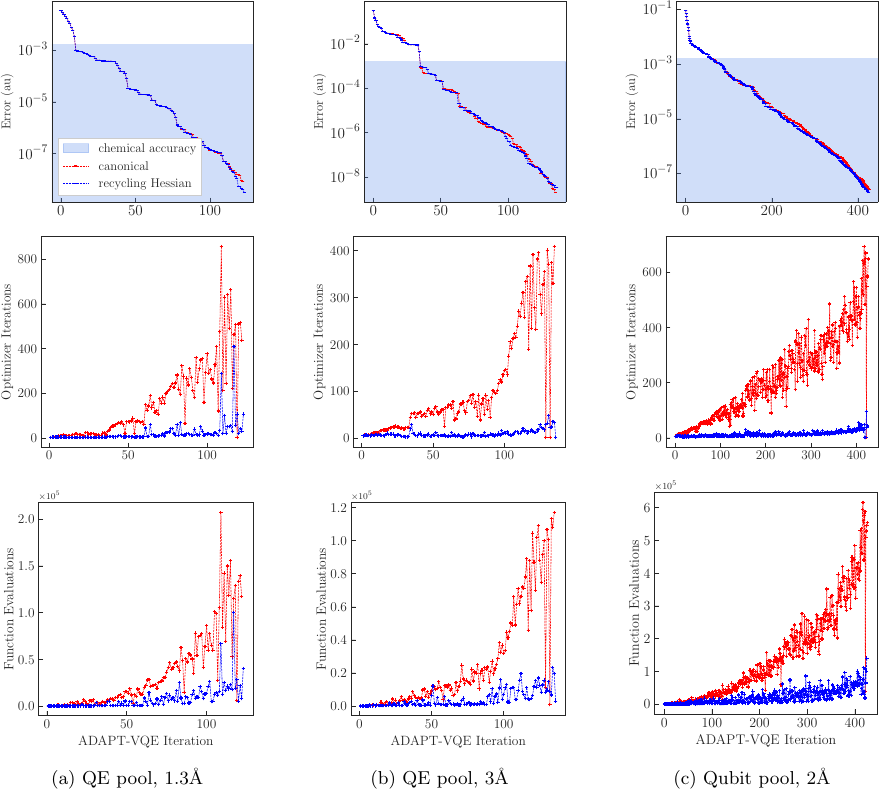}
    \caption{Equivalent plots to Fig.~\ref{fig:h6_results} for BeH$_2$.}
    \label{fig:beh2_results}
\end{figure*}

In the main text, we showed that our Hessian recycling protocol results in a significant decrease in the measurement costs of ADAPT-VQE for H$_6$, and that this decrease tends to become more relevant as the system size increases. In this appendix, we show that these conclusions generalize to other molecules.

Figures \ref{fig:lih_results} and \ref{fig:beh2_results} show the evolution of error, line searches, and measurement costs as the dimension of the ADAPT-VQE optimization grows. The molecules considered are respectively LiH and BeH$_2$. Once again, we consider both equilibrium and stretched geometries and two different pools.

We confirm that recycling the Hessian decreases the total measurement costs to a significant degree for all systems and pools.

\section{Heatmaps for Other Iterations}
\label{ap:extra_hms}

\begin{figure}[htbp]
    
    \includegraphics[width=0.94\columnwidth]{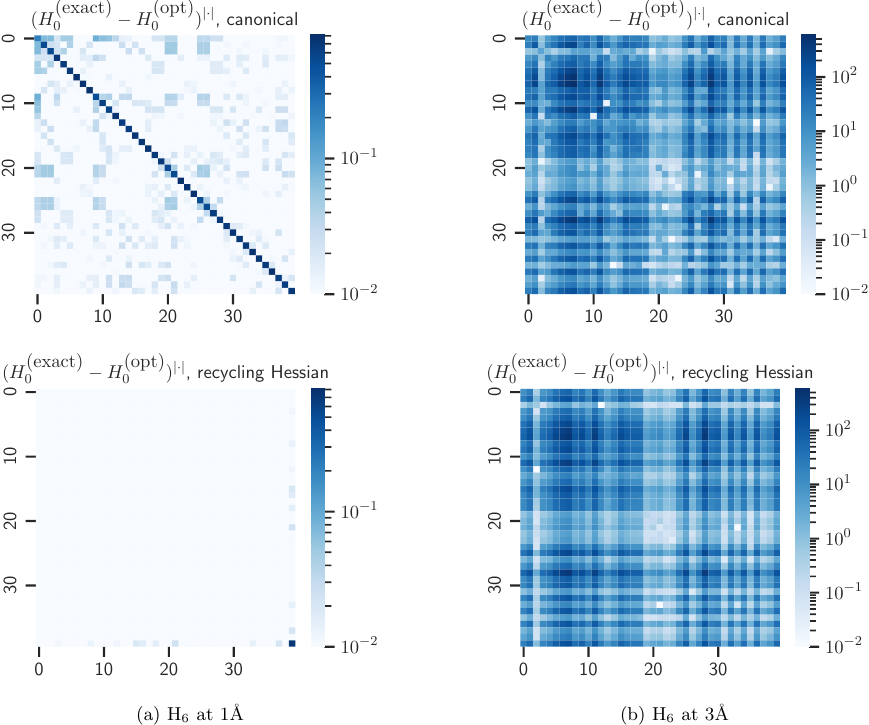}
     \caption{Heatmaps showing the difference between the initial approximate inverse Hessian in the optimization process and the initial exact inverse Hessian, with (bottom row) and without (top row) Hessian recycling, for the 40th iteration of QEB-ADAPT-VQE. The plots show the element-wise difference between these two matrices. We consider H$_6$ at equilibrium and stretched geometries.}
     \label{fig:hm_40}
\end{figure}

\begin{figure}[htbp]
    \includegraphics[width=0.94\columnwidth]{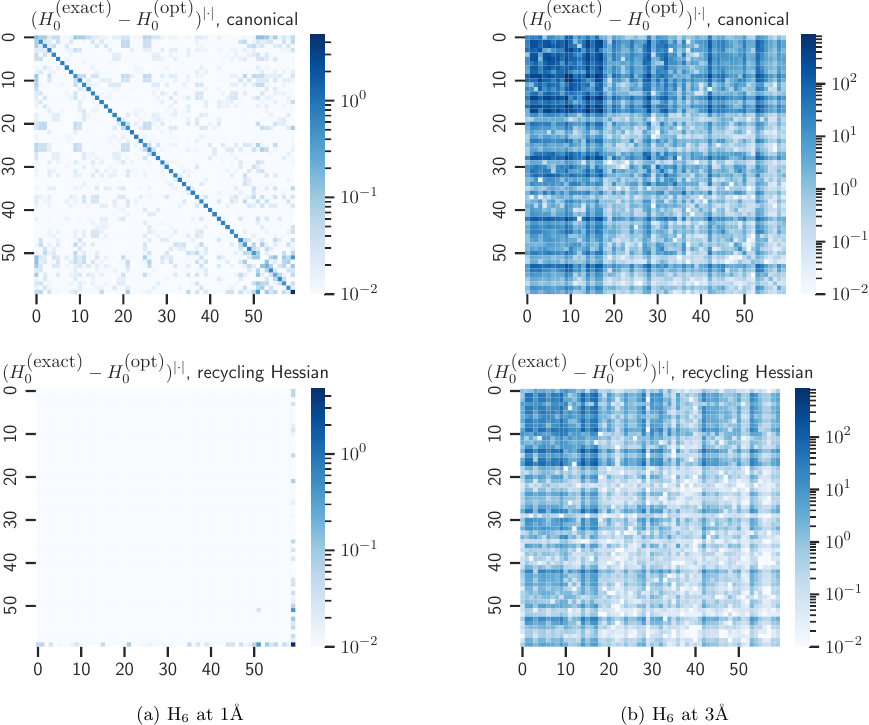}
     \caption{The equivalent of Fig.~\ref{fig:hm_40} for the 60th optimization.}
     \label{fig:hm_60}
\end{figure}

In order to confirm that the behavior showcased in Fig.~\ref{fig:heatmaps} generalizes to other iterations, we consider equivalent heatmaps for other iterations. Figures ~\ref{fig:hm_40} and \ref{fig:hm_60} consider iterations 40 and 60, respectively.
     
As happened in the examples provided in the main text, we observe that the entries of the approximate Hessian are closer to the exact ones when we recycle the Hessian. While this happens for all test cases, once again we see that the impact of our protocol is more significant at the equilibrium geometry. Moreover, as happened in iteration 50, we can observe that the entries farther away from the true value are the diagonal ones for such a configuration, but that is no longer the case for the larger bond distance. The reasons behind these results were discussed in Sec.~\ref{ss:dist_exact}.

\section{Evolution of the 50th QEB-ADAPT-VQE Iteration for H$_6$}
\label{ap:it_50}

\begin{figure*}[htbp]
    \includegraphics[width=\textwidth]{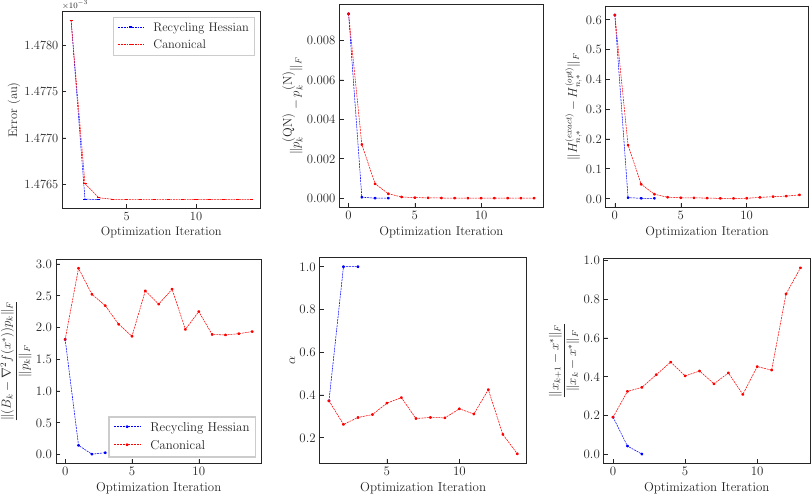}
\caption{Relevant quantities for the study of the convergence rate of the 50th QEB-ADAPT-VQE optimization for H$_6$ at 1\AA{}.}
\label{fig:H6_qe_it50}
\end{figure*}

In this appendix, we analyze a different optimization of QEB-ADAPT-VQE for H$_6$. Our purpose is to confirm that the results observed and discussed concerning the 75th optimization in Sec.~\ref{ss:opt_evol} were not fortuitous.

Figure~\ref{fig:H6_qe_it50} contains the same data depicted in Figs.~\ref{fig:H6_qe_it75} and \ref{fig:H6_qe_it75_convergence}, concerning H$_6$ at 1\AA{}; however, we now focus on the 50th iteration as opposed to the 75th one.

We confirm that the evolution of the optimization is similar. Recycling the Hessian speeds up the convergence of the energy, improves the alignment with Newton's direction, and brings the approximate Hessian closer to the exact one. Additionally, recycling the Hessian allows us to gather all conditions necessary for superlinear convergence, which we verify occurs in the last plot.

\bibliography{main.bib}

\end{document}